# Maximum Chirality in Planar Metasurfaces Induced by Strong Coupling of Quasi-Bound States in the Continuum

Jiaqi Niu,[1,2] Jingquan Liu,[1] and Bin Yang[1,*]

[1]National Key Laboratory of Advanced Micro and Nano Manufacture Technology, Shanghai Jiao Tong university, Shanghai, 200240, China

[2]Department of Micro/Nano Electronics, Shanghai Jiao Tong university, Shanghai, 200240, China

*binyang@sjtu.edu.cn

## ABSTRACT

Achieving intrinsic optical chirality requires breaking all mirror symmetries of an object, and maximum chirality, which allows interaction with only one helicity of light, is particularly promising for applications such as chiral sensing, emission, and lasing. Traditionally, designing maximum chirality in dielectric metasurfaces has relied on precise engineering of vertical symmetry breaking, which presents significant fabrication challenges. Motivated by recent efforts towards enhanced chiral responses in planar structures, we demonstrate that maximum chirality can be achieved in a planar dielectric metasurface through controlled in-plane asymmetries. Specifically, the introduced perturbation induces strong coupling between two accidentally degenerate quasi-bound states in the continuum (QBICs) with orthogonal polarization states, which results in mode splitting into symmetric and antisymmetric modes, each exhibiting opposite circular dichroism (CD) responses. This behavior is quantitatively confirmed using quasinormal mode perturbation theory, by which we also identify a pair of exceptional points (EPs) at the transition between weak and strong coupling regimes. This work expands the existing approaches to maximum chirality in planar structures and aims to inspire future innovations in metasurface design.

## INTRODUCTION

Chirality is a century-old concept that has far-reaching ramifications in various fields such as mathematics, chemistry, biology, and physics. It describes an object that lacks mirror symmetry and therefore is distinguishable from its mirror image. Designing nanostructures with optical chirality is a rapidly evolving research area, owing to the promising applications of chiral imaging, chiral sensing, chiral emission and lasing, as well as other enhanced light-matter interactions with molecules.[1–4] In particular, chiral metasurfaces composed of periodically arranged resonant metaatoms offer unprecedented versatility for the strategic design of various functionalities over a wide range of the electromagnetic spectrum.[5–7] Early investigations of chiral metasurfaces focused on stacked or multilayered structures made of metals,[8–11] which showed a higher degree of optical chirality compared to their planar counterparts as measured by circular dichroism (CD),[12] but they typically suffered from Ohmic loss, thus limiting the quality factors ($Q$). To circumvent this problem, dielectric metasurfaces have gained prominence in recent years due to their negligible loss in the optical spectral regime, and therefore enabling highly resonant chiral features.[13,14] Although breaking all mirror symmetries is essential for intrinsic chirality (nonzero CD without considering cross-polarization effects), it has been found that strong structural chirality does not necessarily lead to strong optical chirality.[15–18] To reach the limit of maximum chirality—that is, the metasurface interacts with only one of the polarization helicities while remaining transparent to the other handedness—two main design principles have been theoretically developed in three-dimensional (3D) structures recently.[19–25] Curiously, both methods are associated with bound states in the continuum (BICs),[26] and consequently, it has been revealed that enhanced CD response is possible even with minor chiral perturbations in the structures.[27]

BICs are resonant states in the limit of infinite quality factor ($Q \to \infty$).[26,28] They are mathematically ideal eigenmodes of an open cavity and can be constructed systematically from symmetry protection or accidentally by destructive interference of several leaky modes. Since BICs are scattering singularities with topological features,[29] and nearby high-$Q$ resonant states (referred to as quasi-BICs, or QBICs) are rich in polarization states,[30,31] it is expected that by strategically perturbing the unit cell structure of a metasurface supporting BICs, practically meaningful QBICs can be designed to exhibit intrinsic chirality (chiral QBICs).[21–24] As mentioned before, there are mainly two established pathways towards chiral QBICs in 3D. The first one is engineering in-plane electric and magnetic dipole (ED and MD) moments by consecutive symmetry-based perturbations in real space.[20–22] It is typically realized by in-plane rotations and out-of-plane offsets/stacking of paired metaatoms in rectangular or elliptical shapes. The second one is manipulating C points in momentum space towards the Γ point (normal direction to the metasurface) by introducing vertical slant/tilt perturbations.[23,24] Both methods have been verified in various experimental setups. However, precise control of the required out-of-plane asymmetry is still challenging for nanofabrication.

It is recognized that enhanced intrinsic chirality is indeed possible in planar dielectric metasurfaces that are compatible with conventional lithography.[32–35] Sharing a similar spirit to the C point tuning method in 3D designs, the slant perturbation is replaced by multilayered metaatoms with composite materials.[33] More intriguingly, managing in-plane ED and MD in planar metasurfaces can be effectively achieved by precisely controlling the thicknesses of dielectric layers, which is easier for practical implementations.[32,34,35] Specifically, enhanced intrinsic chirality has been observed in planar dielectric metasurfaces with fourfold ($C_4$) rotational symmetry,[32,35] in which displacement currents contribute to in-plane MD.[36] The lack of vertical conductive currents in plasmonic metaatoms explains why only weak asymmetry in co-polarized transmission was measured, which is, in fact, due to residual 3D chiral effects, for example, the substrate.[12,37–39] Speaking of the substrate effect, a very recent study showed that the substrate can induce maximum chirality by coupling accidentally degenerate transverse electric (TE)-like and transverse magnetic (TM)-like modes in planar dielectric metasurfaces.[35] The $C_4$ symmetry in these two examples necessitates some loss mechanisms (diffractional and dielectric loss, respectively) required by the reciprocity principle;[32,35] otherwise, CD is zero. By lowering to twofold ($C_2$) symmetry, maximum chirality in a lossless dielectric metasurface is theoretically proposed by perturbing the in-plane symmetry of a single-band TM-like BIC mode.[34]

In this paper, we further explore the approaches towards intrinsic maximum chirality in lossless planar dielectric metasurfaces. In-plane perturbation eliminates all point symmetry elements and strongly couples two accidentally degenerate QBICs, such that the resulting symmetric (or bonding) and anti-symmetric (or anti-bonding) coupled modes correspond to intrinsic chiral states with opposite helicities. The maximum chirality is demonstrated in the transmission spectrum, in which cross-polarization is efficiently suppressed while the metasurface resonantly interacts with only one of the two helicities of normally incident waves. Quasinormal mode perturbation theory (QNMPT)[40] reveals quantitative evidence of the strong coupling, and a pair of exceptional points (EPs) are identified, as should be expected at the transition point from weak to strong coupling regimes.[41] This work is inspired by recent theoretical developments of maximum optical chiralities in planar structures,[34,35] and the unperturbed structure is based on Ref.[42]

## DESIGN AND RESULTS

A metasurface exhibiting maximum chirality should radiate plane waves with pure helicity in its resonant states, or quasinormal modes (QNMs). QNMs are eigenmodes of a leaky cavity, whose eigenvalues are typically complex numbers, with the real parts indicating resonant frequencies and the imaginary parts indicating loss rates. BICs are therefore special QNMs with real eigenvalues.[43,44] As mentioned before, it is typical to tailor maximum chirality at an isolated, nondegenerate QNM.[33,34] However, in this paper, we focus on a more intuitive and systematic approach that couples degenerate TE and TM modes[27,45–47] via symmetry-breaking perturbations, since (1) a plane wave with pure helicity is the sum or difference of phase-matched TE and TM waves, and (2) TE and TM modes are ubiquitous in periodic structures, especially in waveguiding photonic metasurfaces. In addition, TE and TM QBICs can be procedurally constructed from symmetry-protected BICs (SPBICs) according to the selection rule or the Brillouin zone folding (BZF) method.[48,49] As a consequence, weak perturbations (small coupling coefficients) suffice to lead to strong coupling between TE and TM QBICs, since their loss rates and the differences between them are small.[41] This is the reason why maximum chirality is closely associated with QBICs, and small external chiral perturbations can induce strong CD signals.[27] Following this principle, we design a planar metasurface with maximum chirality, enabled by the strong coupling between TE and TM QBICs induced by in-plane structural perturbations.

**Figure 1** depicts the square unit cell structure embedded in vacuum[42], composed of a pair of silicon bars forming a dimer (refractive index $n_{Si} = 3.42$) sitting on a waveguiding layer made of SiO$_2$ ($n_{SiO_2} = 1.45$). The periodicity in the $x$ and $y$ directions is the same ($P = 1000$ nm), and the thicknesses of the SiO$_2$ and Si layers are $t_1 = 230$ nm and $t_2 = 225$ nm, respectively. For the unperturbed structure, the width of the Si bar is $w = 111$ nm. Both bars are aligned along the hatched boundary, and the length of extrusion in the $y$-direction is $L = 520$ nm. These parameters are fixed throughout this paper. As will be discussed later in **Figure 2**, the successive perturbations on the distance between the dimers ($g$), the width of one of the Si bars ($\delta_3$), and the lengths ($\delta_1$ or $\delta_2$) will result in chiral QBICs. Now we focus on a specific set of parameters in **Figure 1** ($\delta_3 w = 50$ nm) to demonstrate the evidence of maximum chiralities in the metasurface. As shown in **Figure 1b**, without chiral perturbation ($\delta_1 = \delta_2 = 0$), the transmittance in the circular basis satisfies $T_{ll} = T_{rr}$ and $T_{rl} = T_{lr}$ due to the remaining in-plane mirror symmetry in the $zx$-plane.[50] Here, $T_{lr} = |t_{lr}|^2$ represents the transmitted energy of left-handed circularly polarized (LCP) waves under right-handed circularly polarized (RCP) incidence. Although the metasurface is geometrically anisotropic in the $x$ and $y$ directions, the accidental degeneracy of two orthogonal QNMs (denoted by M1 and M2 in blue and red color, respectively) near $g = 266$ nm renders it optically almost isotropic. Therefore, the cross-polarizations $T_{rl}$ and $T_{lr}$ are small, which is a prerequisite for maximal chirality. Detailed analysis of this degeneracy can be found in Ref.[42] Here, we introduce additional chiral perturbations $\delta_1$ or $\delta_2$, breaking all point group symmetries. All simulations were performed in COMSOL Multiphysics®.

The results of perturbing the thick bar ($\delta_1 = 0, \delta_2 \neq 0$) are presented in **Figures 1c** and **1d**, and the results for thin bar perturbation ($\delta_1 \neq 0, \delta_2 = 0$) are given in the **Supporting Information (SI)**. We find three notable features in the transmittance spectra in **Figure 1c**. First, the degeneracy in **Figure 1b** is lifted, evidenced by two resonant dips at different incident wavelengths for LCP and RCP, respectively. This indicates that $\delta_2$-perturbation induces strong coupling between M1 and M2. Second, the co-polarization spectra dominate over those of cross-polarization, which means that the metasurface does not change the helicity of the incident waves, in stark contrast to other similar planar designs that rely on extrinsic or false chirality (more details in **Discussion**).[51–53] The reason for such intrinsic chirality roots in the phase difference between M1 and M2, with the substrate contributing to the vertical symmetry breaking (see **Figure 2**). Third, the resonant features can be tuned continuously by sweeping $\delta_2$ from negative to positive values and are almost symmetric with respect to $\delta_2 = 0$. We find that the second-order correction in QNMPT dominates over the first order (more details in **Quasinormal Mode Perturbation Analysis**), which ultimately relates back to the field distributions of M1 and M2. However, this peculiar feature is not found for $\delta_1$-perturbation (see **SI**). **Figure 1d** shows the CD in transmittance derived from **Figure 1c**, which is defined as $CD_T = T_{rr} - T_{ll}$, excluding the polarization conversion effects. Note that our design shares a similar spirit to that reported in Refs.[11,35], but with two major differences. First, it is possible to further boost the $Q$ factors by optimizing structural parameters, since the original modes M1 and M2 are rooted in BICs. Second, our design breaks all point group symmetries; therefore, $CD_T \approx \pm 1$ is possible even if the constituent materials are lossless. In contrast, designs with $C_4$ rotational symmetry must introduce some loss mechanisms; otherwise, the reciprocity relation guarantees $CD_T = 0$.[32]

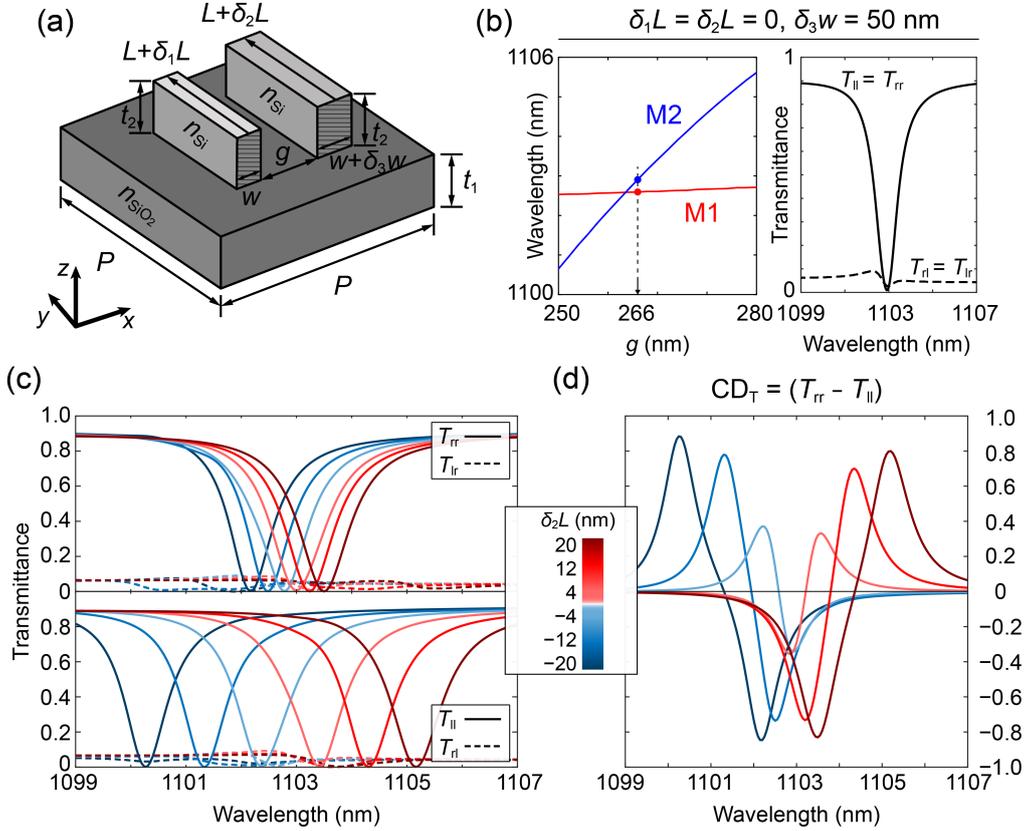

**Figure 1. Unit cell structure and transmittance spectra showing dual maximum chirality.** (a) A pair of silicon bars on a silicon dioxide waveguiding layer, aligned at the hatched surface. Throughout this paper, the following dimensions are fixed: $P = 1000$ nm, $t_1 = 230$ nm, $t_2 = 225$ nm, $w = 111$ nm, and $L = 520$ nm. The refractive indices $n_{Si} = 3.42$ and $n_{SiO_2} = 1.45$ are used in simulations. (b) Left panel: Without chiral perturbation ($\delta_1 = \delta_2 = 0$), mode 1 (M1, $x$-polarized, red) and mode 2 (M2, $y$-polarized, blue) are decoupled due to in-plane mirror symmetry ($\delta_3 w = 50$ nm). An accidental degeneracy (crossing) of M1 and M2 is observed near $g = 266$ nm. Right panel: energy transmittance spectrum ($T = |t|^2$) in the circular basis at $g = 266$ nm; solid lines represent co-polarization, and dashed lines represent cross-polarization. At 1103 nm, the metasurface is effectively optically isotropic ($T_{rl} = T_{lr} = 0$). (c) Transmittance spectra for RCP incidence (upper panel) and LCP incidence (lower panel) for $\delta_2 L = -20, -12, -4, 4, 12, 20$ nm (color-coded from dark blue to dark red). Cross-polarized components (dashed lines) are suppressed compared to co-polarized components (solid lines), indicating near-

maximum chirality. (d) Circular dichroism in transmission ($CD_T$) spectra extracted from (c).

**Figure 2** provides a detailed procedure for designing the metasurface. For modes M1 and M2, the normalized electric fields in the $x$ and $y$ directions, respectively, are plotted, as these components are dominant over others. Note that the normalized QNMs have phases determined up to a sign;[43] therefore, we can directly compare phase shifts in the far field between different modes. The arbitrariness of the sign chosen does not affect the qualitative conclusions, and quantitative comparison between simulation and QNMPT will be discussed in the next section. Starting with $\delta_1 = \delta_2 = \delta_3 = 0$, and with the critical value of $g = g_c = P/2 - w$, mode profiles originating from the BZF are shown in **Figure 2a**. These modes are virtual; however, as $g$ decreases, they evolve into SPBICs (**Figure 2b**). Note that M1 and M2 are associated with modes dominated by electric fields $E_x$ and $E_y$, respectively, on the $zx$ plane cutting through the middle of the dimer. We observe that nodal planes for M1 and M2, denoted by small black arrows in **Figures 2a and 2b**, are located approximately at the central height of the substrate and Si bars, respectively. Upon $\delta_3$-perturbation, both SPBICs turn into QBICs.[54] The wavelengths and amplitudes of the far-field radiations are approximately equal due to near-degeneracy, and the phase shifts are approximately $\pi/2$ in both directions as shown in **Figure 2c**. Such phase shifts are determined by $t_1$ and $t_2$, which are easier to control in nanofabrication compared to slant or tilt in previous designs. Also in **Figure 2c**, the spatially averaged electric fields in the $xy$ plane ($\langle E_{x,y} \rangle_{xy}$) are plotted. Finally, in **Figure 2d**, $\delta_1$- or $\delta_2$-perturbation breaks all mirror symmetries, making the metasurface chiral. As a result, M1 and M2 are strongly coupled, and the symmetric mode (M1 + M2) radiates pure LCP waves, while the antisymmetric mode (M1 − M2) radiates pure RCP waves.[11] Here, we dipict chiral radiation with the time-averaged chiral flux in the $z$ direction ($\langle F_z \rangle$), and the normalized third Stokes parameter $S_3$, which is essentially the average of those calculated in the upper and lower space.[33,55] An equivent depiction of chiral flux, expressed in $E_x \pm iE_y$, is provided in the **SI**. Note that other perturbations, such as rotating the dimers around the $z$-axis, will have similar chiral QBIC effects. We focus on the $\delta_1$-, $\delta_2$-perturbation in this work because it is easier to implement the QNMPT to get quantitative agreement with simulation results.

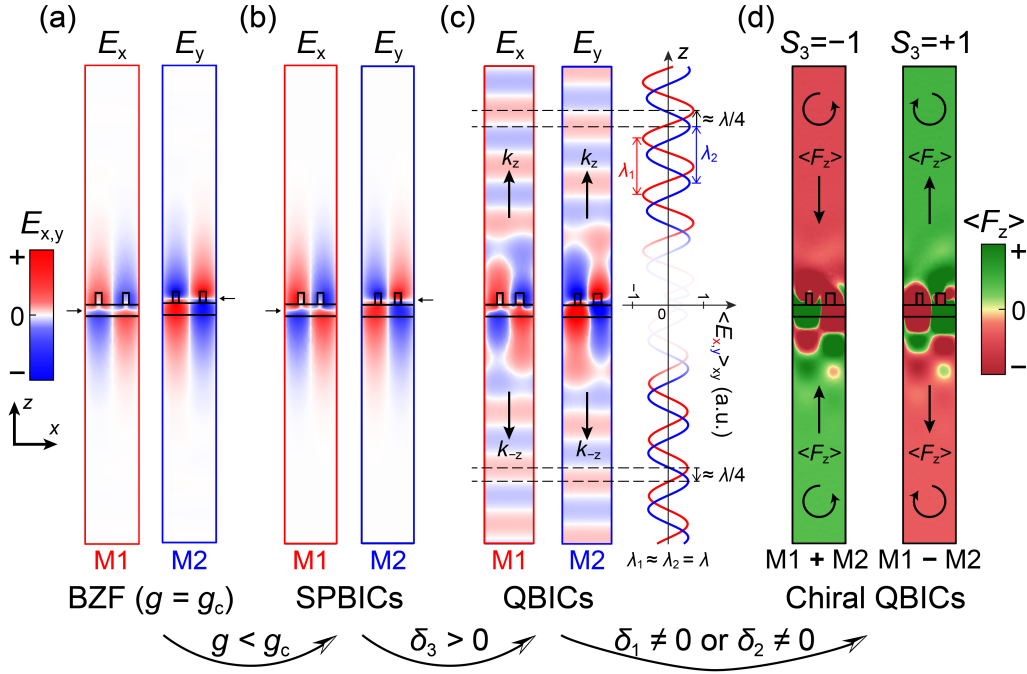

**Figure 2. Qualitative explanation of chiral QBICs in planar metasurface in a successively perturbative manner.** (a) Brillouin zone folding (BZF) at the critical gap distance $g = g_c = P/2 - w$. (b) Reduced gap distance ($g < g_c$) induces symmetry-protected BICs (SPBICs) for both M1 (red) and M2 (blue). The small black arrows in (a) and (b) for M1 (M2) show the nodal plane of the mode profile for $E_x$ ($E_y$), which is the effective symmetry plane for M1 (M2). (c) Quasi-BICs (QBICs) with radiative losses induced by the $\delta_3$-perturbation. Rightmost panel: spatially averaged, normalized electric fields over the unit cell ($\langle \cdot \rangle_{xy}$) in the wave propogating ($\pm z$) direction for M1 ($E_x$, red) and M2 ($E_y$, blue). The wavelengths and amplitudes of M1 and M2 are approximately equal ($\lambda_1 \approx \lambda_2$) because of near-degeneracy (via fine-tuning $g$ and $\delta_3$). The phase difference originates from the vertical displacement of effective symmetry planes in (a) and (b). In (a)–(c), M1 (M2) is tracked by the dominant $E_x$ ($E_y$) components of the normalized QNMs in the $zx$-plane. (d) A pair of chiral QBICs

due to mode coupling and splitting upon $\delta_1$- or $\delta_2$-perturbation. Symmetric (M1 + M2) and antisymmetric (M1 − M2) "dressed" modes are inherently chiral with opposite helicity, characterized by the time-averaged chiral flux ($\langle F_z \rangle$) or equivalently the normalized third Stokes parameter $S_3$.

## QUASINORMAL MODE PERTURBATION ANALYSIS

Quasinormal mode perturbation theory (QNMPT) is a semianalytical method used to predict eigenvalues and field distributions of a perturbed structure, capable of dealing with shape deformations.[40] Technical details have been reported elsewhere,[40] and only major conclusions are reviewed here to suit our purposes. Since only two QNMs, M1 and M2, are relevant in the spectral domain of interest, we suppose the metasurface is a two-level system. The electric fields in the far field for the chiral metasurface ($\delta_2 \neq 0$) can be expressed as the superposition of those of the achiral one ($\delta_1 = \delta_2 = 0, \delta_3 \neq 0$):[40]

$$\mathbf{E}_\pm(\mathbf{r}) = \alpha_1^\pm E_x(\mathbf{r})\hat{x} + \alpha_2^\pm E_y(\mathbf{r})\hat{y}, \qquad (1)$$

where $\alpha_1^\pm$ and $\alpha_2^\pm$ are complex coefficients signifying the proportion of M1 ($E_x(\mathbf{r})\hat{x}$) and M2 ($E_y(\mathbf{r})\hat{y}$) in the perturbed modes, M± ($\mathbf{E}_\pm(\mathbf{r})$). Here, M+ designates the perturbed mode with higher resonant frequency (lower wavelength). Again, we focus on the $\delta_2$-perturbation in the main text and the $\delta_1$-perturbation in the **SI**. According to QNMPT, the eigenfrequencies of the perturbed modes $\omega_\pm$ and the coefficients $\alpha_1^\pm$ and $\alpha_2^\pm$ can be solved from a linear eigenvalue equation, using eigenfrequencies of the unperturbed modes $\omega_1$ and $\omega_2$, and field distributions on the perturbation boundary (colored surface in **Figure S3**; see more details in the **SI**):[40]

$$\begin{pmatrix} \omega_1 & 0 \\ 0 & \omega_2 \end{pmatrix} \begin{pmatrix} \alpha_1^\pm \\ \alpha_2^\pm \end{pmatrix} = \omega_\pm \begin{pmatrix} 1+V_{11} & V_{12} \\ V_{21} & 1+V_{22} \end{pmatrix} \begin{pmatrix} \alpha_1^\pm \\ \alpha_2^\pm \end{pmatrix}. \qquad (2)$$

The analytical expressions for $V_{\beta\gamma}(g;\delta_2)$, where $\beta, \gamma \in \{1,2\}$, are given by Equation S.7.1 in Ref.[40] up to the second order in $\delta_2$.

The second-order perturbation gives better fits for $\delta_2$-perturbation compared to the first-order results, whereas for $\delta_1$-perturbation, the second-order correction is less pronounced (see the **SI**). **Figure 3a and 3b** compare $S_3$ retrieved from the full-wave simulation (left) and second-order QNMPT in **Equation 1** (right). The contour of $|S_3| \approx 1$ traces a parabola-like trajectory, which is even with respect to $\delta_2$. This behavior is not captured by the first-order QNMPT in **Figure S3**. Specifically at $g = 266$ nm, QNMPT gives reasonably good estimates of $S_3$ and resonant wavelengths (**Figure 3c and 3d**), but fails to predict $Q$ at large $|\delta_2 L| > 20$ nm (**Figure 3e**). In this case, the first-order QNMPT gives better reults. **Figure 3f** plots the ratio of the coefficients in **Equation 1** for both M+ (blue) and M− (red). The solid lines and dashed lines are the real and imaginary parts, respectively. We observe that $\alpha_2/\alpha_1$ is almost real; as $\delta_2 \to 0$, $\alpha_1 \to 0$ for M− and $\alpha_2 \to 0$ for M+, consistent with the fact that M+ (M−) reduces back to M1 (M2). In the shadowed region where $|S_3| \approx 1$, $\alpha_2/\alpha_1 \approx \pm 1$ at the same $\delta_2$, which quantitatively validates the arguments in **Figure 2c and 2d**.

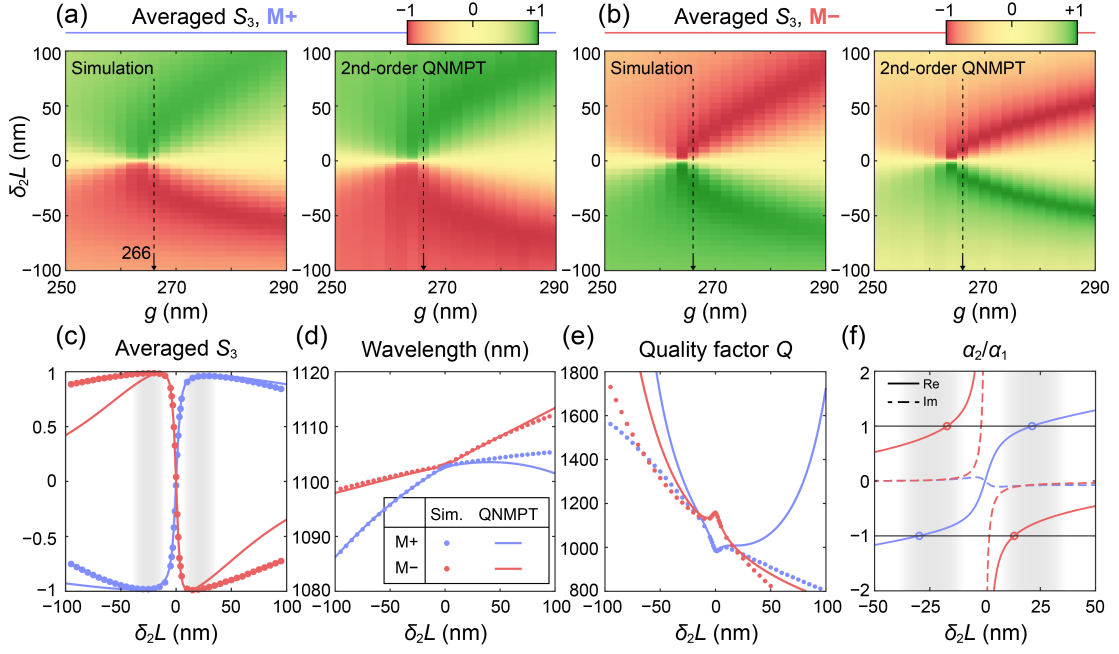

**Figure 3**. Comparison of simulation results and second-order QNMPT fits at $\delta_3 w = 50$ nm. (a, b) Simulation (left) and QNMPT (right) results of the averaged Stokes parameter $S_3$, sampled at upper ($z > 0$) and lower ($z < 0$) radiation channels. (c–e) Comparing simulations (dots) with QNMPT (solid lines) for the coupled modes M+ (blue) and M− (red) at $g = 266$ nm. (c) Averaged $S_3$. (d) Resonant wavelengths. (e) Quality factors. (f) Ratio of expansion coefficients (**Equation 1**), real (solid) and imaginary (dashed) parts. The shadowed area indicates where $|S_3| \approx 1$. Circles highlight the points where $(\alpha_1, \alpha_2) = (1, \pm 1)$ correspond to the symmetric (M1 + M2) and antisymmetric (M1 − M2) coupling, respectively.

## DISCUSSION

An exceptional point is a point of total degeneracy of QNMs in non-Hermitian systems, where both the complex eigenfrequencies and mode profiles coalesce for two or more QNMs.[56,57] EPs robustly exist due to their topological nature, but they are hard to identify because locating them involves tuning multidimensional parameters. Furthermore, it is hard to find exact EPs through brute-force numerical simulations, since they are very sensitive to numerical errors. One possible way to robustly construct EPs is to make use of symmetry.[58] A more general method is to couple two modes, adjusting the coupling strength such that it matches the loss rate difference.[56] In momentum space, it has been found that EPs emerge from the Dirac cone, which is the degeneracy of a SPBIC and a leaky mode.[59] In real space, a recent report shows the coupling of two orthogonal QBICs leads to EPs.[47] Therefore, it is a general feature that EPs can be found by continuously tuning the coupling strength between two originally orthogonal (uncoupled) modes that are degenerate in the real part but have loss differences. Consequently, two EPs are connected by a bulk Fermi arc (BFA), along which the band gap remains closed in the real part.[60]

Equipped with the QNMPT, it is easier to find EPs in our proposed metasurface (**Figure 4a–4c**). Here, we choose $\delta_3 w = 65$ nm in the following analysis. The $V$ matrices in **Equation 2** are numerically calculated for $g \in [290, 310]$ nm, and for each $g$, eigenfrequencies $\omega_\pm$ and the expansion coefficients $\alpha_1^\pm$ and $\alpha_2^\pm$ are obtained by solving **Equation 2** for $\delta_2 L \in [-20, 20]$ nm. **Figure 4a** plots the real and imaginary parts of $\omega_\pm$ in the $g$-$\delta_2 L$ plane, and two EPs denoted by stars are connected by the BFA, as expected. **Figure 4b** shows the phase angle of the eigenvalue difference, $\arg(\omega_+ - \omega_-)$. At EP1 (red star) or EP2 (green star), the complex value of $(\omega_+ - \omega_-)$ equals zero in a topologically protected fashion. The expansion coefficient vectors at EPs in a bilevel system are known to be $(1, \pm i)^T$,[61] which is confirmed in **Figure 4c**. To verify the coalescence of the mode profiles at the EP, numerical simulations are performed with three $\delta_2 L$ values near EP2, as shown in **Figure 4d–4f** at $z = 3t_2/4$ (see **SI** for EP1). It is clear that at EP2, the mode profiles are almost identical (**Figure 4e**). Furthermore, larger $\delta_2$ strongly couples M1 and M2 (**Figure 4d**) into M+ and M− (**Figure 4f**). As can be deduced from **Figure 2c**, the far-field polarizations of EPs are almost linearly polarized, such that the normalized second Stokes parameter $S_2 \approx \pm 1$ (see **SI**).

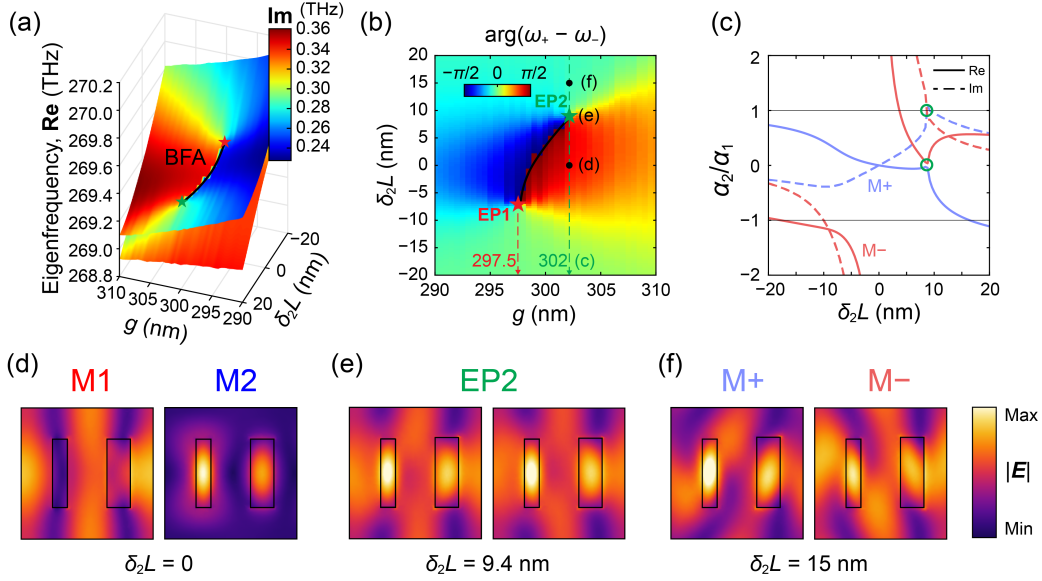

**Figure 4.** Exceptional points (EPs) at the transition point of weak-to-strong coupling regime. (a–c) QNMPT predictions. (a) EPs are highlighted by red and green stars in the Riemann surfaces of complex eigenfrequencies for M+ ($\omega_+$, with larger real part) and M− ($\omega_-$, with smaller real part). EPs are connected by a bulk Fermi arc (BFA, black line), on which $\mathrm{Re}(\omega_+) = \mathrm{Re}(\omega_-)$. (b) Phase angle of $(\omega_+ - \omega_-)$. At EP1 (red star) or EP2 (green star), $\omega_+ = \omega_-$. (c) Ratio of expansion coefficients at $g = 302$ nm. At EP2, $(\alpha_1, \alpha_2) = (1, i)$ indicated by the green circles. (d–f) Full-wave simulations of electric field profiles for two QNMs at the $z = 3t_2/4$ plane at selected points in (b). (d) Decoupled modes at $\delta_2 = 0$. (e) At EP2 with $\delta_2 L = 9.4$ nm, the mode profiles are almost identical, indicating the coalescence of modes at the exceptional point (f) Strongly coupled modes at $\delta_2 L = 15$ nm show significant mixing of the original modes.

This work focuses exclusively on intrinsic maximum chirality, that is, perfect CD without polarization conversion, or equivalently, well-separated QNMs that radiate pure chiral waves with $S_3 = \pm 1$ in the normal direction (at the $\Gamma$ point). It is not trivially realized in planar structures. We differentiate our design from two similar concepts in the literature by comparing the physical mechanisms in detail. First, the extrinsic chirality of planar structures was proposed decades ago,[62,63] in which the up-down mirror symmetry is effectively broken by oblique incidence angles. Later, it was found that such chirality is ubiquitous in planar structures, even with negligible thickness, such as metallic or plasmonic scatterers.[64] From a topological perspective, extrinsic chirality corresponds to C points in momentum space that are split from the V point at the $\Gamma$ point by breaking the rotational symmetry of a SPBIC.[31,65] In contrast, intrinsic chirality requires finite thickness and, additionally, out-of-plane symmetry breaking; otherwise, parity symmetry guarantees $S_3 = 0$.[36] Another form of in-plane chirality, or so-called false chirality, breaks in-plane mirror symmetries only, without involving the vertical dimension.[51–53,66,67] In this case, the reported CD depends on the cross-polarization effect, while for metallic structures with negligible thickness, loss is also necessary. Although it is possible to reach maximal efficiency in polarization conversion due to multiple interference in the cavity, this is not maximum chirality by definition,[19] and its utility is limited for some applications.[23]

We believe that designing planar maximum chirality via mode coupling near accidental degeneracies is a universal strategy. Compared to similar ideas in the literature, our design eliminates the need for loss by breaking the $C_4$ symmetry,[35] and the phase engineering of base modes avoids the tricky search for EPs for enhanced chirality.[47] Therefore, our metasurface also functions as a chiral mirror (helicity-preserving reflection)[68] and is robust over a wide range of structural parameters.

## CONCLUSIONS

In this paper, we propose a method for achieving maximum chirality in planar metasurfaces by synthesizing existing approaches. In particular, we demonstrate a maximally chiral planar metasurface by means of strongly coupling two orthogonally polarized QBICs near an accidental degeneracy. The low-symmetry metasurface is made of a pair of asymmetric dimer bars sitting above a waveguiding layer. Transmittance spectra show suppressed polarization conversion effects, and CD spectra show dual-band chirality of opposite helicity due to mode splitting of strong coupling. The physical origin is clearly illustrated by systematically adding perturbations to the symmetric building block. QNMPT provides quantitative understanding of the mode hybridization process, through which a pair of EPs are easily identified. Approaching intrinsic chirality in planar structures simplifies fabrication processes, and we hope this work will inspire further developments for chiral metasurfaces.

**Supporting Information**

Figure S1 shows results for the perturbation of the thinner bar ($\delta_1$-perturbation). Figure S2 shows far-field radiations of chiral QBICs on the circular basis. Figure S3 shows the first-order QNMPT for $\delta_2$-perturbation. Figure S4 shows the first- and second-order QNMPT for $\delta_1$-perturbation. Figure S5 shows the far-field polarizations, eigenfrequencies, expansion coefficient ratios, and mode profiles at EP1 and EP2.

**Funding**

This work was supported by the National Key Research and Development Program of China (No. 2022YFF1202301), National Natural Science Foundation of China (12072189), the Medicine and Engineering Inter-disciplinary Research Fund of Shanghai Jiao Tong University (No. YG2023QNB25) and the Key Laboratory of Flow Visualization and Measurement Techniques, AVIC Aerodynamics Research Institute (No. XFX20220503). The authors are also grateful to the Center for Advanced Electronic Materials and Devices (AEMD) of Shanghai Jiao Tong University.

**Notes**

The authors declare no competing financial interest.